\documentclass{article}
\usepackage{graphicx,amsmath}
\usepackage{amsfonts}
\DeclareFontFamily{OT1}{rsfs}{} \DeclareFontShape{OT1}{rsfs}{m}{n}{
<-7> rsfs5 <7-10> rsfs7 <10-> rsfs10}{}
\DeclareMathAlphabet{\mycal}{OT1}{rsfs}{m}{n}
\def\scri{{\mycal I}}
\def\scrip{\scri^{+}}%

\begin{document}

\title{Dynamical $SU(2)$  magnetic monopoles}

\author{\textit{Gyula Fodor and Istv{\'a}n R{\'a}cz}\footnote{
Bolyai J\'anos Research Fellow of the Hungarian Academy of Sciences}\\
MTA KFKI, R\'eszecske- \'es Magfizikai Kutat\'oint\'ezet\\
H-1121 Budapest, Konkoly Thege Mikl\'os \'ut 29-33.\\
Hungary\\}

\maketitle

\abstract{
In our paper Phys. Rev. Lett. 92, 151801 (2004), 
the oscillations of strongly deformed
Bogomolny-Prasad-Sommerfield magnetic monopoles have been studied 
on a fixed Minkowski background.
The purpose of the present article is to provide a more detailed account on 
the results yielded by our numerical simulations.
In particular, an analysis on the dependence on the strength of the initial 
excitation is carried out in order to distinguish features which are already 
present for small perturbations from those which are consequences of the 
nonlinearity of the system.
}

\section{Introduction}  \label{intro}

The investigation of solitons in particle physics is of fundamental
interest (see e.g. \cite{rub} for a recent review). In particular,
considerable attention  has been paid to the  study of
't\,Hooft-Polyakov magnetic monopole solutions of coupled
Yang-Mills--Higgs  (YMH) systems \cite{tH,P}.
Magnetic monopoles are important because they are present in a wide range 
of gauge theory models. 
Moreover, their existence can be used to provide an account on various 
physical phenomena, including e.g.\ the quantized nature of electric charge.
In a  recent paper by the present authors \cite{ymhprl} a report is provided 
on the numerical study on the behavior of strongly excited 
Bogomolny-Prasad-Sommerfield (BPS) magnetic monopoles \cite{bo,ps}.
In a related article by Forg\'acs and Volkov \cite{fv} the same system is 
investigated by perturbative methods.
There is a striking consistency between the key results of these two 
independent studies.
The aim of the present paper is to provide more detailed description of
the system based on the results obtained by 
the numerical investigations of \cite{ymhprl}.

The investigated dynamical magnetic monopole is a solution of a
coupled $SU(2)$ YMH system. The Yang-Mills field is represented by an
$\mathfrak{su}(2)$-valued vector potential $A_a$ and the associated 2-form
field $F_{ab}$ reads as
\begin{equation}
F_{ab}=\nabla_aA_b-\nabla_bA_a+i g\left[A_a,A_b\right]\label{ymf}
\end{equation}
where $[\ ,\ ]$ denotes the product in $\mathfrak{su}(2)$ and $g$
stands for the gauge coupling constant. The Higgs field (in the
adjoint representation) is given by an $\mathfrak{su}(2)$-valued
function $\psi$ while its gauge covariant derivative reads as
$\mathcal{D}_a\psi= \nabla_a\psi+i g[A_a,\psi]$. The dynamics of the
investigated YMH system is determined by the Lagrangian
\begin{equation}
\mathcal{L}=Tr(F_{ef}F^{ef})
+2Tr(\mathcal{D}_e\psi\mathcal{D}^e\psi)
+\lambda\left[Tr(\psi\psi)-v^2\right]^2,
\end{equation}
where $\lambda$ is the self interaction constant of the Higgs field.
In order to have finite energy solutions
the value of $Tr(\psi\psi)$ must tend to $v^2$ as the distance from the 
monopole goes to infinity.
The energy-momentum tensor of the YMH field takes the form
\begin{equation}
T_{ab}=-\frac{1}{4\pi}\left[Tr(F_{ae}{F_b}^{e})
-Tr(\mathcal{D}_a\psi\mathcal{D}_b\psi)
+\frac{1}{4}g_{ab}\mathcal{L}\right].
\end{equation}

Our considerations are restricted to  spherically symmetric
configurations yielded by the `minimal' dynamical generalization of
the static 't\,Hooft-Polyakov magnetic monopole configurations
\cite{tH,P} (see also \cite{Hua}). Accordingly, the evolution takes
place on Minkowski spacetime the line element of which, in spherical
coordinates $(t,r,\theta,\phi)$, is
\begin{equation}
ds^{2}=dt^{2}-dr^{2}-r^{2}\left(  d\theta^{2}+\sin^{2}\theta\,d\phi
^{2}\right),
\label{ds}
\end{equation}
while the Yang-Mills and Higgs fields, in the so called {\it abelian
gauge}, are assumed to posses the form
\begin{eqnarray}
\hskip -.25cm A_a=-\frac{1}{g}\left[w\left\{\tau_{_{2}}(d\theta)_a-
\tau_{_{1}}\sin\theta(d\phi)_a \right\} +
\tau_{_{3}}\cos\theta(d\phi)_a\right] \\
\psi=H \tau_{_{3}},\phantom{+\left[\tau_{_{3}}\cos\theta(d\phi)_a\right]}
\end{eqnarray}
where the generators $\{\tau_{_{I}}\}$ (I=1,2,3) of $\mathfrak{su}(2)$
are related to the Pauli matrices $\sigma_{_{I}}$ as $\tau_{_{I}}=
\frac12\sigma_{_{I}}$, moreover, $w$ and $H$ are assumed to be smooth
functions of $t$ and $r$.

In the Bogomolny-Prasad-Sommerfield (BPS) limit, i.\ e. when
$\lambda\rightarrow0$, the mass of the Higgs boson becomes much less 
than the mass of the vector boson.
The field equations in the $\lambda=0$ case become
\begin{eqnarray}
& & r^{2}{\partial ^{2}_r}{w}- r^{2}{\partial ^{2}_t}{w} =
w\left[\left({w}^{2}-1\right)+g^2{r^{2}}{H}^2 \right]
\label{ymhe22}  \\
& &\hskip 0.35cm r^{2}{\partial ^{2}_r}{H}+ 2r{\partial_r }{H}
- r^{2}{\partial ^{2}_t}H =
2{w}^{2}H \ , \label{ymhe11}
\end{eqnarray}
with the restriction that
\begin{equation}
\lim_{r\rightarrow\infty}H=v \label{Hinf}\ ,
\end{equation}
in order to represent the BPS limit of finite energy solutions.
We note that the time independence of the limit value of $H$ at 
both spacelike and null infinity follows from the field equations
(\ref{ymhe22}) and (\ref{ymhe11}) (see \cite{fr2}).

The system of equations
(\ref{ymhe22}) and (\ref{ymhe11}) with condition (\ref{Hinf}) 
has an analytic solution, the static BPS monopole  \cite{bo,ps}
\begin{equation}
w_0 = \frac{gv r}{{\rm sinh}(gvr)}\ ,\ \ \ \ \ \ \ \
H_0 = v\left[\frac{1}{{\rm tanh}(gvr)}-\frac{1}{gv r}\right]\ .\label{BPS}
\end{equation}
In the rest of the paper considerations will be restricted to the 
$\lambda = 0$ case and strong impulse type excitations of the BPS 
monopole will be investigated.

In the BPS limit the Higgs field becomes massless and the only scale
parameter of the system is the vector boson mass $m_w=gv$.
Since in the case considered here $v\not=0$, the rescalings
$t\rightarrow \tilde t=t m_w$, $r \rightarrow \tilde r=r m_w$ and
$H\rightarrow \tilde H=H/v$ transform the parameters to the value
$g=v=1$.
This implies that it suffices to consider the evolution of the
$g=v=1$ system numerically and to study merely the dependence of 
the evolution on the various initial conditions for this system. 

To have a computational grid covering the full physical spacetime --
ensuring thereby that the outer grid boundary will not have an
effect on the time evolution -- the technique of conformal
compactification, along with the hyperboloidal initial value problem,
is used. This way it is possible to study the asymptotic
behavior of the fields close to future null infinity, as well as, the
inner region for considerably long physical time intervals.

The conformal transformation we use is a slight modification of
the static hyperboloidal conformal transformation applied by Moncrief
\cite{mon}. It is defined by introducing first the new
coordinates $T$ and $R$ instead of $t$ and $r$ as
\begin{equation}
T=\omega t+1-\sqrt{\omega^2 r^2+1}
\ \ \ {\rm  and} \ \ \  
R=\frac{\sqrt{\omega^2 r^2+1}-1}{\omega r},
\end{equation}
where $\omega$ is an arbitrary positive constant. 
The Minkowski
spacetime is covered by the coordinate domain given by the
inequalities  $-\infty < T < +\infty $ and $0 \leq R < 1$.
The $R=const.$ lines (same as the $r=const.$ lines) represent 
world-lines of `static observers', while the $T=const.$ 
hypersurfaces are hyperboloids of the Minkowski spacetime satisfying 
the relation
$(\omega t+1-T)^2-\omega^2 r^2=1$.
The constant $\omega$ is introduced 
here in order to have control on the size of the monopole region
in terms of the $R$ coordinate, which is useful for the optimization 
of the numerical code. In the rest of the paper we present results
obtained by setting $\omega=0.05$.

The line element of the conformally rescaled metric 
$\widetilde{g}_{ab}=\Omega^2{g}_{ab}$ 
in the coordinates $(T,R,\theta,\phi)$ takes the form  
\begin{equation}
d\widetilde s^2=\frac{\Omega^2}{\omega^2}dT^2+2RdTdR-dR^2-
R^2\left(d\theta^2+{\rm sin}^2\theta\,d\phi^2\right) 
\end{equation}
where the conformal factor is
\begin{equation}
\Omega={\omega}(1-R^2)/{2}\ .
\end{equation}
The $R=1$ coordinate line represents
$\scri^+$ through which the conformally rescaled metric
$\widetilde{g}_{ab}$ smoothly extends
to the unphysical coordinate domain with $R>1$.
The simple relation $r\Omega=R$ holds between the old and new radial 
coordinates.

Using the substitution
$H(t,r)={h(t,r)}/{r}+v$ 
the field equations (\ref{ymhe22}) and (\ref{ymhe11}) in the new
coordinates read as
\begin{eqnarray}
&& {\mathfrak P}{ w}
=  w\left[\left({ w}^{2}-1\right)+
g^2\left( h+v
R\Omega^{-1}\right)^2\right] \label{2ymhe24}\\ &&
\phantom{{\mathfrak P}{\mathfrak P}{\widetilde w}}  
{\mathfrak P}{ h}=
2\left({h}+v{R}{\Omega}^{-1} \right) w^{2},
\label{2ymhe14}   
\end{eqnarray}
where the differential operator $\mathfrak P$ is defined as 
\begin{eqnarray}
\hskip -.35cm {\mathfrak P} &=&
\frac{4R^2}{(R^2+1)^2}\left[\frac{\Omega^2}{\omega^2}{\partial 
^{2}_R} - {\partial ^{2}_T} - 
2R{\partial_R\partial_T}\right. \nonumber \\ & &
\phantom{\frac{4R^2}{(R^2+)}} 
\left.-\frac{2\Omega}{\omega(R^2+1)}{\partial_T} -
\frac{\Omega R(R^2+3)}{\omega(R^2+1)} 
{\partial_R}\right]. 
\end{eqnarray}
These equations can be put into the form of a first order strongly
hyperbolic system \cite{fr2}. The initial value problem for such a
system is known to be well-posed \cite{gko}. In particular, we solved
this first order system 
numerically by making  use of the `method of line' in a fourth
order Runge-Kutta scheme following the recipes proposed by Gustafsson
{\it et al} \cite{gko}. All the details related to the numerical
approach, including representations of derivatives, treatment
of the grid boundaries are to be published in
\cite{fr2}. The convergence tests justified that our code 
provides a fourth order representation of the selected evolution
equations. Moreover, the monitoring of the energy conservation and the
preservation of the constraint equations, along with the coincidence
between the field values which can be deduced by making use of the
Green's function and by the adaptation of our numerical code to the
case of massive Klein-Gordon fields, made it apparent that the
phenomena described below have to be, in fact, physical properties of the
magnetic monopoles.

In each of the numerical simulations initial data on the $T=0$
hypersurface was specified for the system of our first order
evolution equations. In particular, a superposition of the 
data associated with the BPS monopole, (\ref{BPS}),
and of an additional pulse of the form 
\begin{equation} \hskip -.27cm(\partial_T w)_\circ = \left\{
\begin{array} {rr}  
    c \exp\left[\frac{d}{(r-a)^2-b^2}\right], & {\rm if}\ r\in
    [a-b,a+b] ;\\[4mm]
  0 \  , & {\rm otherwise},  \end{array} \right.\label{ff}
\end{equation}
with $a \geq b>0$, which is a smooth function of compact support, was
used. This 
choice, providing non-zero time derivative for $w$, corresponds
to ``hitting'' the static monopole configuration 
between two concentric shells at $r=a-b$ and $r=a+b$ with a bell shape
distribution.  Basically the same type of evolution occurs when
instead  of  $(\partial_T w)_\circ$ we prescribe $(\partial_T
h)_\circ$ in a similar fashion. 
The only distinction is that in this
later case the energy of the pulse is given directly to the massless
Higgs field which carries immediately a considerably large portion
of it to $\scri^+$ before the Yang-Mills field could take it over to
feed it into the breathing state of the monopole and into the expanding
oscillations of the massive Yang-Mills field.

All the simulations shown below refer to the
same pulse shape (\ref{ff}) corresponding to the choice of the parameters 
$a=2$, $b=1.5$ and $d=10$, only the amplitude $c$ is varied.
It is important to keep in mind that 
the energy of the pulse is proportional to $c^2$. 
For example, for the case $c=70$ presented in \cite{ymhprl} it is 
$55.4\%$ of the energy of the static monopole, which is clearly not
just a simple perturbation. 
We would
also like to emphasize that the figures shown below are typical
in the sense that for a wide range of the parameters characterizing
the exciting pulse, qualitatively, and in certain cases even
quantitatively, the same type of responses are produced by the
monopole \cite{fr2}.

\begin{figure}[!htb]
\begin{center}
      \includegraphics[ scale=1.4]{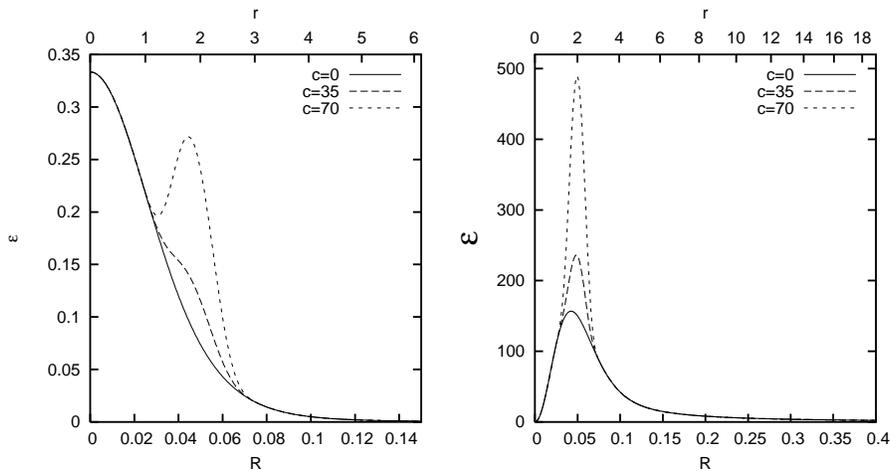}
      \caption{Energy density distributions $\varepsilon$ and ${\mathcal{E}}$
      on the $T=0$ initial hypersurface for the static BPS monopole ($c=0$) and
      for two different choices of the initial amplitude $c$.}
 \label{figindat}
\end{center}
\end{figure}

In order to provide a better picture about the size and distribution of the
monopole and its relation to the initial data it is instructive to plot
the energy density distribution on the $T=0$ initial hypersurface.
The conserved energy current vector $j^a=T^a_{\ \ b}u^a$ is formed from the 
energy momentum tensor $T^a_{\ \ b}$, where $u=\frac{\partial}{\partial t}$ 
is the
velocity vector of the the static observers characterized by the constant 
value of $r$ along their worldline.
The projection $\varepsilon=j^a n_a
=\omega T^0_{\ \ 0}(R^2+1)\Omega^{-1}/2$,
where $n_a$ is the future pointed
unit normal to the constant $T$ hypersurfaces, gives a gauge invariant
quantity characterizing the energy density on these surfaces.
The energy ``density'' associated to shells of radius $R$
can be calculated as
${\mathcal{E}}=\int\varepsilon\sqrt{|h|} d\theta d\phi
=2\pi\omega T^0_{\ \ 0}(R^2+1)R^2\Omega^{-4}$,
where $h$ is the determinant of the
induced metric on the constant $T$ hypersurfaces.
On Fig.\ \ref{figindat} we plot the value of $\varepsilon$ and ${\mathcal{E}}$
for the static monopole both with and without an initial deformation of the
form of (\ref{ff}).
The plot of ${\mathcal{E}}$ is more instructive because its integral from
the center $R=0$ to null infinity $R=1$ (i.\ e.\ the area below the curve)
gives the total energy of the system along the chosen constant $T$ 
hypersurface.
As the time passes, the value of this integral decreases
exactly by the energy radiated to null infinity.

\begin{figure}[!thb]
\begin{center}
      \includegraphics[ scale=1.1]{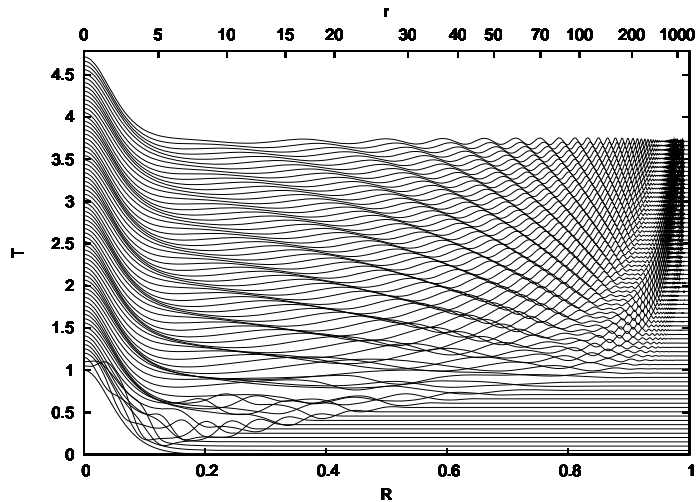}
      \caption{Spacetime diagram showing the
 time evolution of the variable $w$ corresponding to the massive
 Yang-Mills component of the gauge field for the time interval
 $0\leq T\leq 3.71$. The amplitude of the oscillations can be evaluated
 by observing that at the center $w=1$, while at infinity $w=0$, always
 holds.
} \label{figw}
\end{center} 
\end{figure}

\begin{figure}[!htb]
\begin{center}
      \includegraphics[ scale=1.1]{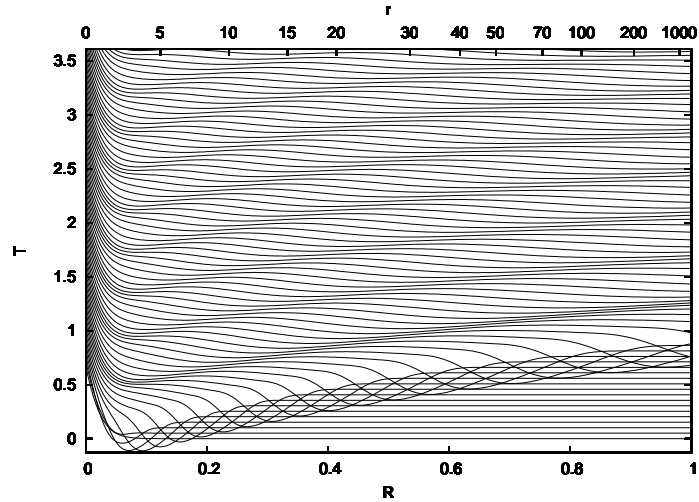}
      \caption{Spacetime diagram showing the
 time evolution of the variable $h$ corresponding to the massless Higgs
 component of the gauge field for the initial amplitude choice $c=70$.
 At the center $h=0$ for all time,
 while at infinity $h$ remains $-1$ until the radiation arrives there.
} \label{figh}
\end{center}
\end{figure}

Choosing temporarily $c=70$, we consider next the quantity $w$,
corresponding to the massive Yang-Mills field, plotted 
on succeeding $T=const.$ hypersurfaces, providing thereby a
spacetime picture of its time evolution (see Fig.\,\ref{figw}). 
One can see the formation of expanding shells of high frequency
oscillations in the asymptotic region \cite{fr}. 
These shells take out energy
from the central monopole region, although they never reach null
infinity because of their massive character.
The time evolution of the other gauge variable, the massless Higgs
field $h$ is strikingly different (see Fig.\,\ref{figh}).
There is no sign of the high frequency oscillations in $h$, justifying
that the two fields effectively decouple in the asymptotic region.
Since the oscillations in $h$ are nonzero at null infinity, the direct
energy transport to $\scri^+$ by the Higgs field, with the velocity of 
light, is apparent.

\begin{figure}[!htb]
\begin{center}
      \includegraphics[ scale=1.1]{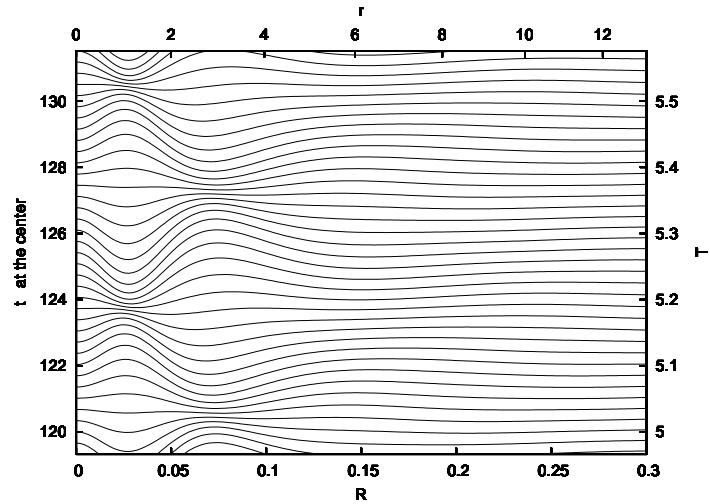}
      \caption{Spacetime diagram showing the
      time evolution of the energy density difference
      ${\mathcal{E}}-{\mathcal{E}}_0$ from $T=4.97$ to
      $T=5.58$. } \label{figt00sdm}
\end{center}
\end{figure}

From Fig.\,\ref{figw} and Fig.\,\ref{figh} it appears that the lower
frequency oscillations are not concentrated only at the monopole but
they are present basically everywhere.
However, if one considers the clearly physically
meaningful quantity, the energy density ${\mathcal{E}}$
corresponding to spherical shells of radius $R$,
it is apparent that the
oscillations are concentrated at the central region,
forming a long lasting `breathing state' of the monopole.
In Fig. \ref{figt00sdm} we show the energy density difference
${\mathcal{E}}-{\mathcal{E}}_0$ in a large central region, where
 ${\mathcal{E}}_0$ is the energy density of the static monopole.
(See Fig.\ 1 of\ \cite{ymhprl} for the plot of ${\mathcal{E}}$ up 
to null infinity.)
Although Fig.\ \ref{figt00sdm} covers a relatively early time period, the
picture remains essentially the same for the rest of the time evolution
when taking time intervals of the same length, with the only difference
of an overall decrease of its amplitude proportional to $T^{-5/6}$
(see Fig.\ 5 of \cite{ymhprl}).

\begin{figure}[!htb]
\begin{center}
      \includegraphics[ scale=1.1]{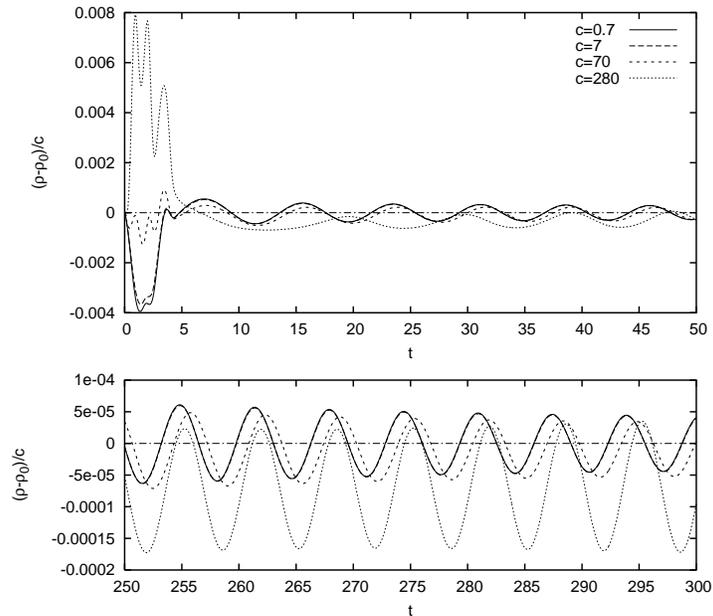}
      \caption{Evolution of the rescaled energy density difference 
      $(\rho-\rho_0)/c$ as seen by a constant radius observer 
      at $R=0.0254$ (i.\ e.\ $r=1.018$) for four different values of $c$ and
      two different time intervals. 
} \label{figeps}
\end{center}
\end{figure}

Next we consider the dependence of the energy density oscillations
on the amplitude of the initial deformation.
The energy density $\rho=T_{ab}u^au^b$ observed by static Minkowski
observers cannot be used to form a conserved quantity when integrating
on the hyperbolic $T=const.$ hypersurfaces.
However, for our specific choice $\omega=0.05$ the $T=const.$
hyperboloids are so close to the flat $t=const.$ surfaces in the inner
monopole region that plots of the quantities $\varepsilon$ and $\rho$ 
are almost indistinguishable.
Furthermore, for the static BPS solution $\varepsilon=\rho\equiv\rho_0$.
On Fig.\ \ref{figeps} we plot the energy density fluctuation
$\rho-\rho_0$ around the static monopole value $\rho_0$ for four
different choices of the amplitude $c$.
Since for small initial perturbations
the deviations of the density $\rho$ from the static value $\rho_0$ 
appear to be
proportional to the amplitude of the perturbation, we rescale the amplitude 
of the four graphs accordingly, by plotting the value 
$(\rho-\rho_0)/c$.

For $c=0.7$ and $c=7$ the corresponding curves almost entirely
overlap, showing that these excitations are still in the linear
domain.
Indeed, the energy provided by the initial perturbation in those two
cases are $0.00554$ and $0.554$ percent, respectively, of the
energy $4\pi$ of the static monopole.
For the $c=70$ case the energy given to the system is $6.966$, which
is $55.4$ percent of the energy of the initially static monopole.
In this case the behavior is clearly nonlinear initially, but after
the direct pulses have left the system the quasinormal oscillation of the
monopole is still not too far from the linear regime.
For $c=280$ the additional energy is $111.46$, which is
more than $8$ times the energy of the static monopole.
In this case even the monopole oscillations show a different character
at the beginning, having a significantly lower frequency.
However, in all cases, the frequency of the monopole oscillations tend
as $t^{-2/3}$ to the asymptotic value $1$ corresponding to the 
vector boson mass in our rescaled coordinates
(see Fig.\ 5 of \cite{ymhprl}). 

After the initial period, i.e.\ for $t>5$, the monopole starts a 
quasi-periodic oscillation with amplitude decreasing as $t^{-5/6}$.
It is visible that the mean value of these oscillations, which represents 
the average energy density at the given radius is negative for large 
initial deformations.
This means that the energy contained in
the oscillating monopole region is actually smaller than the energy of
the static monopole which is a clear manifestation of the 
nonlinear character of the evolution.  
As it was pointed out to us by Michael Volkov, this situation is similar 
to that where a large stone is dropped into a lake, and the water level 
decreases because of the submerging of the stone. 
This analogy, however, does not provide a completely satisfactory
explanation in the monopole case, because the 
average energy remains negative during the entire evolution, approaching 
merely in the asymptotic limit the static value from below.
This might lead to the interpretation that the monopole behaves like an
extremely viscosal fluid.

\begin{figure}[!htb]
\begin{center}
      \includegraphics[ scale=1.1]{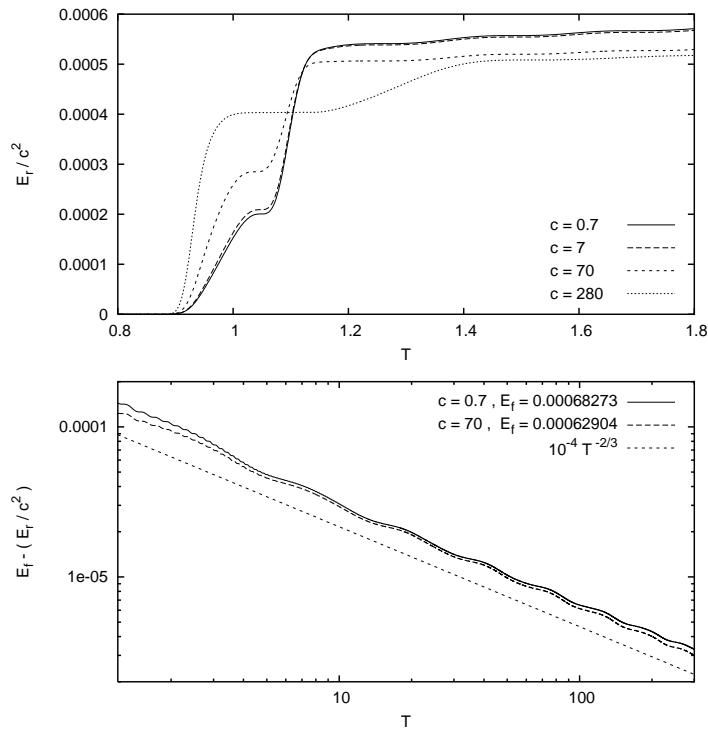}
      \caption{The rescaled energy $E_r/c^2$ radiated by the system
until time $T$ for various initial amplitudes $c$. On the lower figure
the logarithmic plot of the difference from the asymptotic value $E_f$
is displayed for a longer time period, indicating that the energy tends to
$E_f$ as $T^{-2/3}$. 
} \label{figsint}
\end{center}
\end{figure}

In the last part of the paper we consider the energy radiated to null 
infinity up to the time $T$.
This energy can be calculated as $E_r=\int_0^T SdT$ where 
$S=\lim_{r\rightarrow\infty}j^a k_a$ with $k_a$ being the unit normal to
the constant $r$ hypersurfaces.
The possibility that we could study this radiation, in particular the 
evolution near $\scrip$, is due to the fact that the hyperboloidal initial 
value problem was used in our numerical simulations.
The radiation arrives to $\scrip$ when the light cone emanating from the 
outer edge of the initial perturbation reaches null infinity, 
i.e.\ approximately
at $T=0.86$.
For small perturbations the radiated energy is proportional to the
square of the initial amplitude. 
Hence we plot the rescaled energy $E_r/c^2$ on Fig.\ref{figsint}
for various initial data specifications.

The energy given to the system by the initial deformation is
approximately $0.0014216\cdot c^2$. 
For small initial perturbations the total radiated energy is
$0.00068273\cdot c^2$, which is $48.02$ percent of the energy given to the 
system.
The rest of the energy is in the self-similarly expanding shells formed
by the massive Yang-Mills field $w$.
These shells get arbitrarily far from the monopole region, but because
of their massive character their
velocity is smaller than the speed of light, whence they never reach null
infinity.
Actually, the energy in the shells is initially a little bit larger than the 
remaining $51.98$ percent, since as we have seen, the average energy in 
the central monopole region is smaller than the energy of the 
static monopole solution in the same region.
As the time passes the amplitude of the monopole oscillations decreases
and the mean value of the energy density approaches the static value from 
below.
Accordingly, there is, although a low scale, energy transfer back to the 
monopole from the outer regions ensuring, in turn, that in the inner 
region the system will settle down to the static monopole solution.

We would also like to emphasize that -- as it is intuitively expected --
the ratio between the energy carried by the shells and the energy 
radiated to $\scrip$ depends on the profile of the initial
excitation. 
It is found that in case of the specific choice (\ref{ff}) this ratio 
depends on $a$, $b$ and $d$, 
but for small amplitudes in the linear regime it is apparently 
independent of the parameter $c$.
 
Various new features of the evolution of excited $SU(2)$ BPS monopoles 
have been presented here. 
The results contained in the last part of this paper indicate that there
remained a lot of interesting issues to be studied both numerically and
analytically in the dynamical behavior of the t'Hooft-Polyakov monopoles.

{\bf Acknowledgments} The authors wish to thank P\'{e}ter
Forg\'{a}cs who suggested the numerical investigation of magnetic
monopoles.
We also would like to thank him, Michael Volkov and J\"{o}rg
Frauendiener for discussions.
This research was supported in parts by OTKA
grant T034337, TS044665 and NATO grant PST.CLG. 978726.

\vfill\eject
\end{document}